\documentclass[a4paper,11pt]{article}
\usepackage{pos}

\usepackage{adjustbox}
\usepackage{amsmath}
\usepackage{graphicx}
\usepackage{booktabs}
\usepackage{subfigure}

\title{
Gluonic gravitational form factors of the proton }

\author*[a]{Zein-Eddine Meziani}

\affiliation[a]{Argonne National Laboratory,\\
 9700 South Cass Avenue Lemont, IL 60439, USA}

\emailAdd{zmeziani@anl.gov}

\abstract{The gravitational form factors (GFFs) are a fundamental and elegant way to describe the structure of nucleons and nuclei. Their Fourier transform allow a description of the spatial distribution of the mass, angular momentum, pressure and shear force densities for both quarks and gluons in the nucleon. While previous investigations predominantly focused on the proton electromagnetic form factors (EMFFs) leading to the charge and magnetization distributions determination, the current emphasis has shifted towards expanding our understanding of the gravitational form factors of quarks and gluons where little is known. In particular, more recently, the proton {\it gluonic} GFFs have been the target of an intensive investigation at Jefferson Lab. This endeavor,  is not without its challenges, particularly in navigating the complexities associated with the near-threshold region. Nevertheless, it provides a bedrock for future nucleon and nuclei gluonic structure studies at the future EIC. In this talk, I will focus on the recent results of $J/\psi$ photoproduction near-threshold on the proton at Jefferson Lab to determine, in particular, the elusive {\it gluonic} gravitational form factors. We discuss the caveats of their extraction in the threshold region and mention the complementary measurements of $\Upsilon$ at the EIC critical to access the trace anomaly and gain insight into the origin of the nucleon mass.}

\FullConference{25th International Spin Physics Symposium (SPIN 2023)\\
 24-29 September 2023\\
 Durham, NC, USA\\}

\begin{document}
\maketitle

\section{Introduction}
The mechanical properties of the proton are best described by the energy-momentum tensor form factors~\cite{Kobzarev:1962wt,Pagels:1966zza}. In relation to the higher three-dimensional parton (transverse position and longitudinal momentum) distributions describing the structure of the nucleon, these form factors can be derived from the generalized parton distributions (GPDs) once an integration over the parton longitudinal momentum fraction $x$ is performed, and a Frourier transform over the transverse position is carried-out. Since these form factors are investigated using tensor (i.e.,  graviton-like ($J^{PC}= 2^{++}$ ) or scalar ($J^{PC} = 0^{++}$) exchange probes, they inherited the name "{\it gravitational}" form factors (GFFs). A comprehensive discussion of the proton GFFs is provided in \cite{Burkert:2023wzr} and in these proceedings \cite{Lorce:2024ipy} with an emphasis on the quark sector where more is known.  However, gluons are the bosons responsible for the confinement of quarks and themselves into a color neutral proton and provide a large fraction of its mass, thus, in their own right, they deserve as much attention as quarks. Here, we are particularly interested in probing the elusive, electrically neutral but color charged gluons confined in a proton.  This task comes with its own experimental challenges due to the lack of a direct gluons' probe as well as theoretical challenges for interpreting the measurement using indirect probes. In this contribution we describe the experimental measurements of the photoproduction of $J/\psi$ in the near-threshold region as a possible path to access some of the GFFs. We present a first determination of two of the four gluonic GFFs using the measured doubly  differential ( in photon energy $E_{\gamma}$ and 4-momentum transfer $t$)  photoproduction cross sections, and  compare them to those calculated using the most potent {\it ab initio} calculations,  those of lattice QCD. We further use the GFFs Fourier transform in the light-front frame according to~\cite{Lorce:2018egm,Freese:2021czn} where a quasi-probabilistic interpretation in term of densities of the mechanical properties is warranted. Finally, we show some of these mechanical properties, like mass, pressure, and shear forces densities as a function of $r_{\perp}$, the transverse distance from the center of the moving proton in the longitudinal direction. We consider these results as the start of an important path aiming at mapping the gluonic structure of the proton through its GFFs. These results should be confirmed or refuted once the gluon GPDs using $J/\psi$ production at large center of mass photon-nucleon energy are determined at the electron-ion collider\cite{Accardi:2012qut}. Another avenue to confirm or refute these results is to use a heavier quarkonium, in this case the $\Upsilon$  and measure its photo- or electro-production near-threshold at the electron-ion collider\cite{Gryniuk:2020mlh,AbdulKhalek:2021gbh} and extract hopefully the same gluonic form factors but with more controlled approximations. 

\section{The experiments at Jefferson Lab}

The GlueX experiment in Hall D observed, for the first time, near-threshold $J/\psi$ production off a proton target at Jefferson Lab. The collaboration  published their measurement of a differential photoproduction cross section at one average photon beam energy of 10.7 GeV as well as the total photoproduction cross section in the near-threshold region in  ref.~\cite{GlueX:2019mkq} .  The data of the differential cross section was used in ref.~\cite{Kharzeev:2021qkd} to provide a mass radius by combining all GFFs into one effective GFF in the near-threshold region, and a  vector meson dominance model was used to describe the photoproduction cross section. Since their first $J/\psi$ publication GlueX collected and analyzed more data which they published more recently,  they include differential cross sections in three different incident photon energy bins, ranging  8.20 GeV $ < E_{\gamma} <$ 9.28 GeV, 9.28 $< E_{\gamma} <$  10.36 GeV and 10.36 GeV $< E_{\gamma} <$ 11.44 GeV, respectively,  with $|t|$ values as high as 8 GeV$^2$ for the highest energy bin~\cite{GlueX:2023pev}.

The Jefferson Lab experiment E12-16-007~\cite{Meziani:2016lhg} also known as $J/\psi-007$ was performed in Hall C at Jefferson Lab and consisted of using the CEBAF electron beam at an energy of 10.6 GeV passing through an 8.5\% copper radiator upstream of a central pivot of rotation of two high momentum spectrometers (HMS \& SHMS). The mixed beam of electrons and photons  passed through a 15 cm liquid hydrogen target located at this pivot producing $J/\psi$ particles. The $J/\psi$ e$^+$e$^-$ pair decays were detected in the HMS and SHMS respectively while the recoil proton was not detected. Nevertheless estimates of the incoherent background, where an extra pion was produced, as well as the Bethe-Heitler contamination were found to be negligible and consistent with our description of the invariant mass of the lepton pair spectrum identifying the $J/\psi$ peak. After detectors efficiency corrections for each spectrometer, and combined spectrometers acceptance corrections at different kinematics settings, the doubly-differential cross sections where unfolded in two-dimensional bins of different  photon energy $E_{\gamma}$ and four momentum transfer to the proton $t$. The description of the experiment and resulting differential cross sections are shown in ref.~\cite{Duran:2022xag} in comparison with several cross section models~\cite{Kharzeev:2021qkd,Mamo:2019mka,Guo:2021ibg,Hatta:2018ina,Hatta:2019lxo,Sun:2021gmi} .

Finally, in Hall B other near-threshold photoproduction experiments were performed using the CLAS12 detector. The analysis of the collected data is underway and the collaboration is expecting to finalize the results in the summer of 2024~\cite{CLAS12}.

\section{Gluonic gravitational form factors}
The matrix element of the QCD energy-momentum tensor for quarks or gluons separately reads~\cite{Ji:1996ek}:
\begin{align}
&\langle p_f, s_f | T_{q,g}^{\mu,\nu}(0) | p_i,s_i \rangle = \\  \nonumber
&\bar u(p_f,s_f) \displaystyle{\Bigl ( } A_{q,g}(t) \gamma^{ \{ \mu} P^{\nu \} }  + B_{q,g} \frac{iP^{ \{ \mu }\sigma^{\nu \} \rho}\Delta_{\rho}}{2M_N} 
+ C_{g,q} \frac{\Delta^{\mu}\Delta^{\nu} -g^{\mu\nu}\Delta^2}{M_N} + \bar C_{q,g}(t) M_N g^{\mu,\nu} \displaystyle{ \Bigr )} u(p_i,s_i)
\end{align}

\noindent where $(p_i , s_i )$ and $(p_f, s_f )$ correspond to the momentum and polarization of the incoming
and outgoing nucleon, respectively, $P= (p_i + p_f )/2, = p_f- p_i, t = \Delta^2$. $A_{q,g}(t)$, $B_{q,g}(t)$, $C_{q,g}(t)$, and $\bar C_{q,g}(t)$ are the quarks and gluons GFFs of the nucleon. It is worth noting that $ J(t)= 1/2 [ A(t)+B(t) ]$ where $J_{q+g}(0)=1/2$ is the total spin of the proton. Furthermore, it is known from lattice calculations~\cite{Pefkou:2021fni,Hackett:2023rif} that $B_{g}(t)$ is small and roughly flat across $t$,  thus it was dropped from the expression of cross sections. Finally,  when added $\bar C_{q+g}=\bar C_{q}(t) + \bar C_{g}(t) = 0$ does not contribute to the matrix element of the energy momentum tensor. Since $\bar C_g(t)$ is not known except at $t=0$~\cite{Hatta:2018sqd,Tanaka:2018nae,Tanaka:2022wzy} we had no choice but to ignore it in our extration at this time. However, its impact should be examined rigorously in future studies. Because, we know that $\bar C(0)$ is sensitive  to the trace anomaly of the energy momentum tensor~\cite{Hatta:2018sqd,Tanaka:2018nae,Tanaka:2022wzy}, its effect must be important and cannot be totally ignored. Last but not least, $C_{q,g}$ form form factors  sometimes  are labeled $D_{q,g}$  with the relation $D_{q,g}(t) = 4C_{q,g}(t)$. The $D(t)$ form factor was dubbed the "Druck (pressure)-term", it is also found in the expansion of the GPDs~\cite{Polyakov:1999gs,Pasquini:2014vua}.

\begin{figure}[t]
\includegraphics[width=\textwidth]{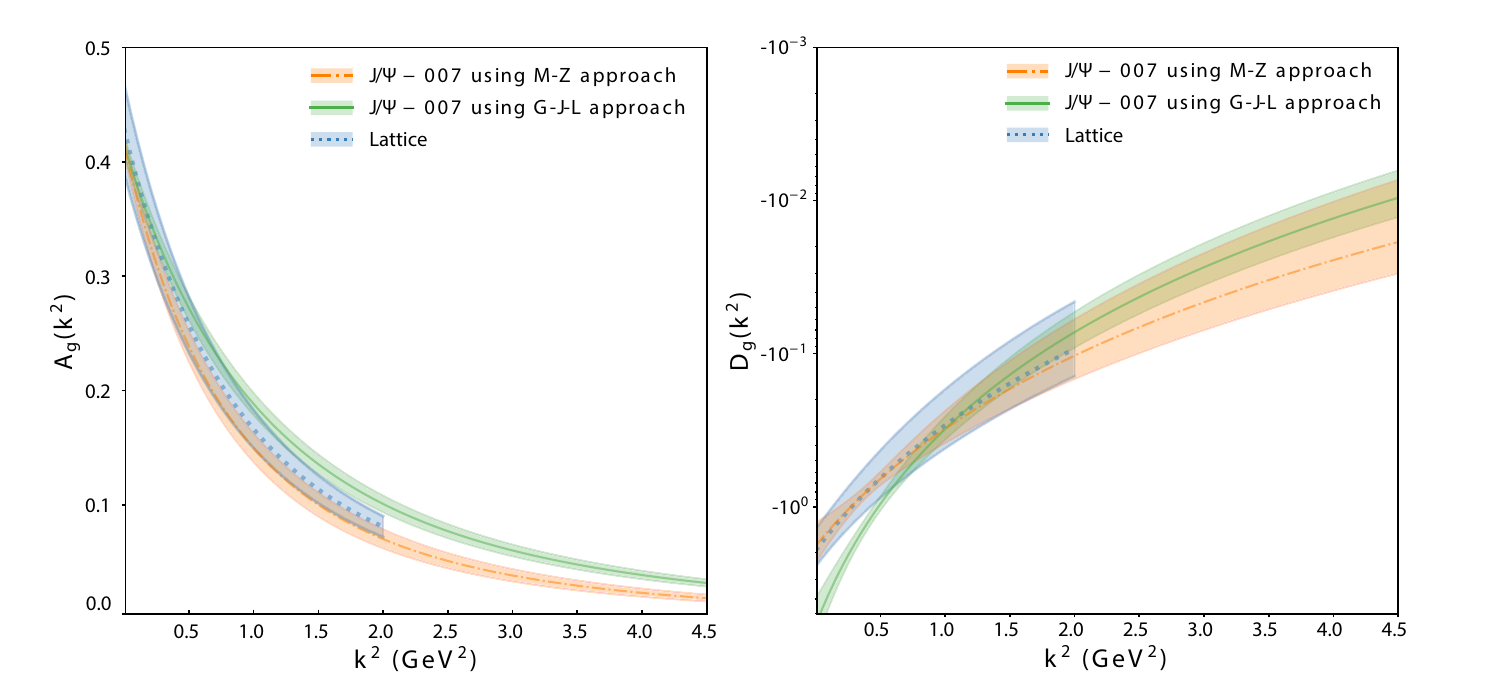}
\caption{
Left panel: The $A_g(k^2)$ form factor ($k^2=|t|$) extracted from our two-dimensional cross section data in the holographic QCD approach~\cite{Mamo:2019mka,Mamo:2021krl} (orange dash-dot curve) and in the updated GPD approach~\cite{Guo:2023pqw} (green solid curve), compared to the latest lattice calculation~\cite{Pefkou:2021fni,Hackett:2023rif} (blue dotted curve). All form factors used the tripole approximation form. The shaded areas show the corresponding uncertainty bands. Right panel: The extracted $D_g(k^2)=4C_g(k^2)$ form factor with the same color scheme as the left panel.}
\label{fig:gff-ac}
\end{figure}

\section{Models of Photoproduction Cross Sections For GFFs Extraction}
Two-dimensional fits of the cross sections with two very different models were performed. One a holographic QCD model by Mamo and Zahed (referred to as M-Z)~\cite{Mamo:2019mka,Mamo:2022eui} and a GPD model by Guo, Ji and Liu (referred to as G-J-L) \cite{Guo:2021ibg},  superseded by an updated calculation by Guo, Ji, Liu and Yang~\cite{Guo:2023pqw}.

The expressions of the cross section for the two different models we used, namely M-Z and G-J-L, to perform the two-dimensional ($E_{\gamma}$ and $t$) fit and extract both the $A_g(t)$ and $C_g(t)$ form factors. In the M-Z holographic model the cross section is expressed as follows:
\begin{equation}
\frac{d\sigma}{dt} = {\cal N} \times \frac{e^2}{64\pi(s-M^2)^2}\times \frac{[ A_g(t) + \eta^2 D_g(t) ]^2}{A_g^2(0)}
\end{equation}

$A(t)$ and $D(t)$ shapes are fully calculated in this model (see \cite{Mamo:2022eui}), however, a dipole or tripole  functional form is assumed, as a very good approximation, for both form factors and was used to fit the data. The normalization constant  ${\cal N}$ = 7.768 GeV$^{-4}$ is taken from ~\cite{Mamo:2019mka}. For a consistent comparison with the lattice QCD calculations of ref.~\cite{Pefkou:2021fni,Hackett:2023rif} we chose in our fitting procedure of the differential cross sections a tripole functional form for both $A_g(t)$ and $C_g(t)$ form factors,  
\begin{eqnarray}
A_g(t) = A_g(0)/\left ( 1- \frac{t}{m_A^2} \right )^3,~~~~~~~~~C_g(t) = C_g(0)/\left ( 1- \frac{t}{m_C^2} \right )^3
\end{eqnarray}
Where $A(0)$ is the average momentum fraction carried by the gluons in the nucleon and determined by the CT18 global parametrization~\cite{Hou:2019efy} of the world DIS data, $\langle x \rangle_g =0.414\pm 0.008$. The three other parameters, namely $m_A$, $C(0)$ and $m_C$ are free parameters determined by the two-dimensional fitting of the differential cross section data of the $J/\psi -007$ experiment.

In the Generalized Parton Distribution (GPD) approach of J-G-L ~\cite{Guo:2021ibg,Guo:2023pqw}, the cross section is expressed as follows:
\begin{equation}
\frac{d\sigma}{dt}= \frac{\alpha_e m e^2_Q}{4(s-M^2)^2} \frac{(16\pi\alpha_s)^2}{3M^2_V}\vert \psi_{NR} \vert^2 
\vert G(t,\xi) \vert^2 
\label{eqn:xsecgpd}
\end{equation}
where $M_{V}$ is the mass of the vector meson, in this case the mass of the $J/\psi$. $\psi_{NR}(0)$ is related to the non-relativistic wave function of the $J/\psi$ at the origin, $G(t,\xi)$ contains the proton gluons GPDs and $e_Q$ is the charge of the quark in the unit of proton charge.
\begin{equation}
\vert G(t,\xi) \vert^2 = \frac{4}{\xi^4} \{ \bigl (1-\frac{t}{4M_N^2}\bigr ) E_2^2 -2E_2(H_2+E_2) +(1-\xi^2)(H_2+E_2)^2 \},
\label{eqn:amplisquared}
\end{equation}
\begin{equation}
\int_0^1dx H_g(x,\xi,t)  =  A^g_{2,0}(t) + (2\xi)^2C_2^g\equiv H_2(t,\xi), 
\int_0^1dx E_g(x,\xi,t)  =  B^g_{2,0}(t) - (2\xi)^2C_2^g\equiv E_2(t,\xi)
\end{equation}

In this approach we have also chosen a tripole functional form for the GFFs embedded in the expressions of $H_2$ and $E_2$. Once $m_A$, $C(0)$ and $m_C$  are obtained $A_g(t)$ and $C_g(t)$ are determined. The latter are shown in Fig.~\ref{fig:gff-ac} and used to extract the mass and scalar radius (\ref{eqn:m_radius}) for each model (see Table~\ref{jpsi:fitparams-one}). Please note that the values of the form factors and corresponding radii using the G-J-L model and presented here have been updated according to ~\cite{Guo:2023pqw}. The mass (scalar) radius is expressed as
\begin{align}
    \langle r_{m,(s)}^2\rangle_g  = \left.6 \frac{1}{A_g(0)}\frac{dA_g(t)}{dt} \right\vert _{t=0} - 6 (18) \frac{1}{A_g(0)} \frac{C_g(0)}{M^2_N}
    =\frac{18}{m_A^2} - 6 (18)\frac{1}{A_g(0)} \frac{C_g(0)}{M^2_N}
\label{eqn:m_radius}
\end{align}

\begin{table*}[ht!]
\caption{ The gluonic GFFs fit parameters, proton mass radius and scalar radius 
determined from the $J/\psi-007$ experiment~\cite{Prasad:2024} through a two-dimensional fit using the holographic QCD approach~\cite{Mamo:2019mka,Mamo:2021krl,Mamo:2022eui} and updated GPD approach~\cite{Guo:2023pqw}. The corresponding proton mass and scalar radii are also shown according to eq.~(\ref{eqn:m_radius}).
In all cases we used the tripole-tripole functional form approximation for the GFFs. We compare these results to the lattice calculations~\cite{Pefkou:2021fni,Hackett:2023rif}.}
\begin{adjustbox}{width=\linewidth}
\begin{tabular}{ccccccc}
\toprule
Theoretical approach & $\chi^2$/n.d.f &$m_A$ (GeV) &  $m_C$ (GeV) & $C_g(0)$ &$\sqrt{\langle r_m^2\rangle}_g$ (fm) & $\sqrt{\langle r_s^2\rangle}_g$ (fm)\\
\midrule
Holographic QCD  & 0.925 &1.575$\pm$0.059 & 1.12$\pm$0.21 & -0.45$\pm$0.132 & 0.755$\pm$0.067 & 1.069$\pm$0.126\\
\midrule
GPD & 0.986 &1.825$\pm$0.063 & 0.774$\pm$ 0.092 & -1.478 $\pm$ 0.467 &   1.074   $\pm$ 0.144&   1.744   $\pm$ 0.259  \\
\midrule
Lattice & & 1.641$\pm$ 0.043 & 1.07$\pm$ 0.12 & -0.483$\pm$ 0.133 & 0.7464$\pm$0.055 &1.073$\pm$0.114 \\
\bottomrule
\end{tabular}
\end{adjustbox}
\label{jpsi:fitparams-one}
\end{table*}

\begin{figure}[b]
\includegraphics[width=\textwidth]{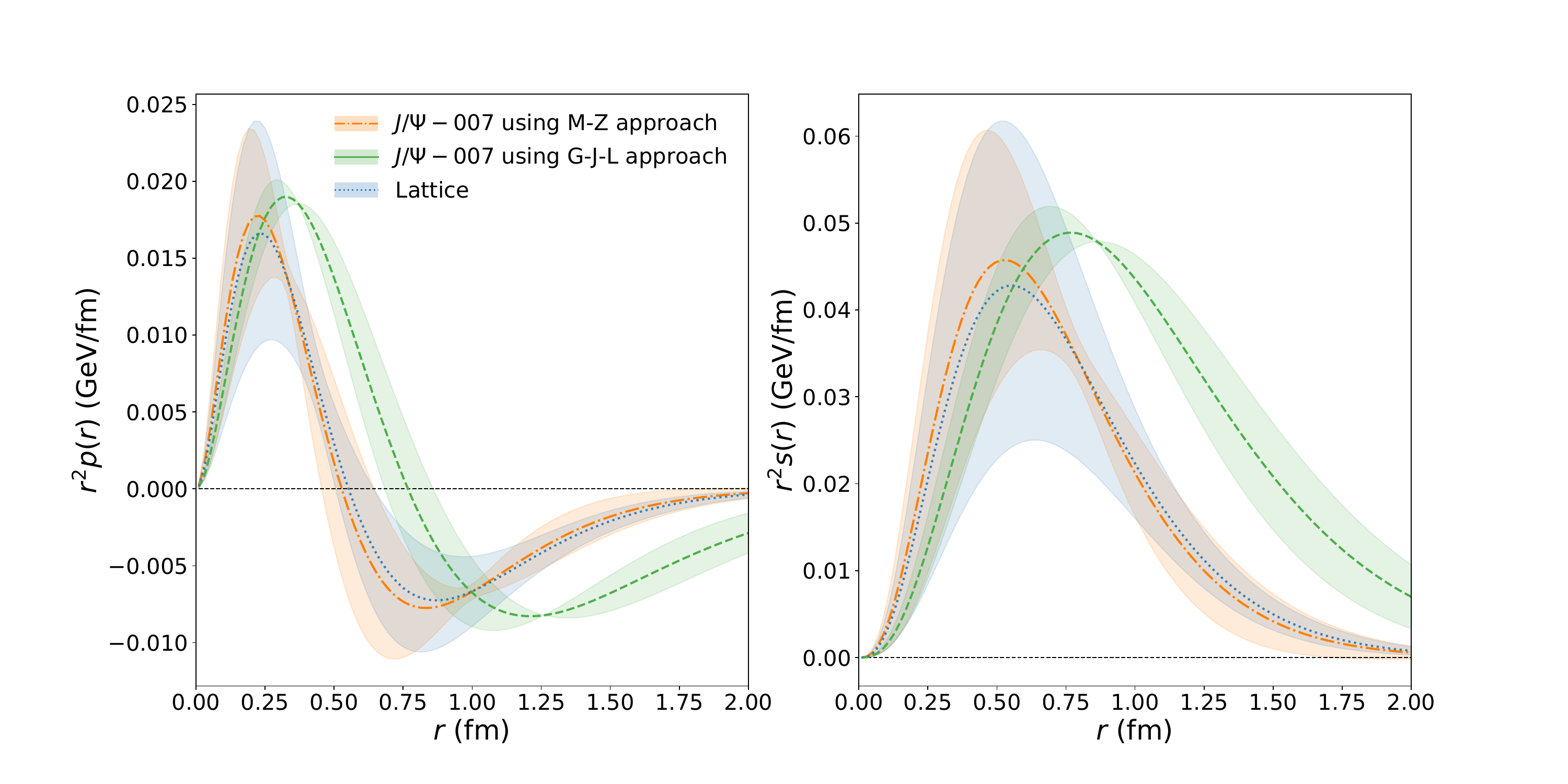}
\caption{Left panel: The $r^2p_g(r)$ pressure in the Breit frame in the holographic QCD approach~\cite{Mamo:2019mka,Mamo:2021krl,Mamo:2022eui} (orange dash-dot curve) and in the updated GPD approach~\cite{Guo:2023pqw} (green solid curve), compared to the latest lattice calculation~\cite{Pefkou:2021fni,Hackett:2023rif} (blue dotted curve). The shaded areas show the corresponding uncertainty bands. Right panel: The extracted shear forces density $r^2s_g(r)$ of the gluons in the same Breit frame and color scheme as the left panel.}
\label{fig:ft-ps}
\end{figure}
We note that the holographic QCD approach gives results close to the lattice QCD calculations~\cite{Pefkou:2021fni,Hackett:2023rif} . For the GPD approach the $A_g(t)$ form factor is compatible with both the holographic QCD extraction and the lattice calculation. However, a difference outside the uncertainties is noticed when $t$ is close to zero for the $D_g(t)$ form factor, which, of course, has an impact on the determination of radii (mass and scalar) as well as the extracted mechanical properties.

To perform our fitting and define the radii listed in Table~\ref{jpsi:fitparams-one} we have neglected the GFF $B_g(t)$ and ignored $\bar C_g(t)$. The holographic QCD approach is a non-perturbative approach that seems to agree well with lattice QCD, although the lattice calculations are still performed away from the pion physical mass (m$_{\pi}$=400 MeV). The GPD approach is a perturbation method and has been used at leading order in an expansion of $1/\xi$ where $\xi$ has to be very close to 1. The results point to the importance of investigating the higher order terms since the bulk of the data are not that close to unity. It appears that both methods give similar values for the $A_g(t)$ GFF but differ for $D_g(t)$ GFF especially at low $t$ where $\xi$ is not close to unity. Of course, it is premature to favor one approach versus the other at this stage to extract the gluonic GFFs in the vicinity of the threshold region. Further measurements in the full $\xi$ phase space, especially close to $\xi = 1$ dominating the large $t$ region, are required to better understand the approximations of the GPD method, for example. In the case of the holographic QCD method more precise data in the photon energy range between 11GeV and 20 GeV would be very valuable to test the energy dependence. Such measurements of $J/\psi$ electro and photo-production on the proton have been proposed and approved at Jefferson Lab using the SoLID~\cite{JeffersonLabSoLID:2022iod} while measurements of both electro and photo-production using a smaller dipole size, namely $\Upsilon$, have part of the EIC science program at the EIC~\cite{AbdulKhalek:2021gbh,Gryniuk:2020mlh}  
\section{Gluon mechanical properties in the Breit frame}
In order to obtain the gluon contribution to the proton mechanical properties, namely the pressure and shear forces densities as a function of distance $r$, we  performed a Fourier transform of the $D(t)=4C(t)$ according to: 
\begin{equation}
    \Tilde{D}(r) = \int \frac{d^3\Delta}{(2\pi)^3}e^{-i\Delta\cdot r}D(\Delta,m_C)=\int \frac{d^3\Delta}{(2\pi)^3}e^{-i\Delta\cdot r}\frac{D(0)}{(1+\frac{\Delta^2}{m_C^2})^3}=D(0)\frac{m_C^3}{32\pi}(1+m_Cr)e^{-m_Cr}
\end{equation}
with  $-t=k^2=\Delta^2$.  The pressure $r^2p(r)$ and the shear forces $r^2s(r)$ densities in the Breit frame are given by~\cite{Polyakov:2018zvc}:
\begin{equation}
    r^2p(r)=\frac{1}{6m_N}\frac{d}{dr}\left (r^2 \frac{d}{dr}\Tilde{D}(r) \right),~~~
    r^2s(r)= -\frac{1}{4m_N}r^3\frac{d}{dr}\left (\frac{1}{r}\frac{d}{dr}\Tilde{D}(r)\right )
\end{equation}
Here we used the tripole functional form for the $C_g(t)$ form factor in both models, consistent with the lattice calculations result. The results~\cite{Prasad:2024} of pressure and shear forces densities in the Breit frame as a function of distance $r$ are shown in Fig.~\ref{fig:ft-ps}. However, since we know that in the Breit frame the interpretation of these Fourier transforms in terms of densities is questionable for fast recoiling targets, like in our case, we decided to also show the results below in the light-front frame.

\begin{figure}[b]
\includegraphics[width=\textwidth]{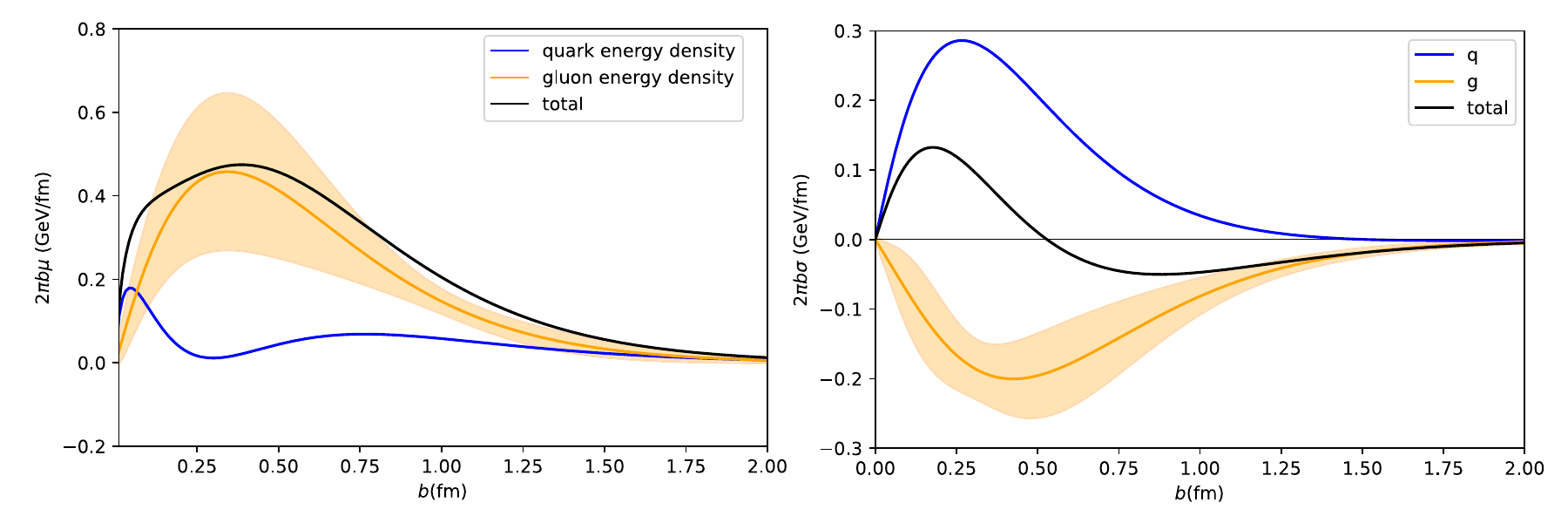}
\caption{Left panel: The $2\pi b \mu_g(r)$ energy density in the light front according to ~\cite{Lorce:2018egm} with GFFs extracted using the holographic QCD approach~\cite{Mamo:2019mka,Mamo:2021krl,Mamo:2022eui} (orange dash-dot curve). The shaded area shows the corresponding uncertainty band for the gluons case only. Right panel: The extracted pressure density $2\pi b \sigma_g(r)$ of the gluons in the light front frame with the same color scheme as the left panel.}
\label{fig:ft-mup}
\end{figure}

\section{Gluons mechanical properties in the light-front frame}
Similar to the electromagnetic case, it has been argued that the better frame to define true densities which correspond to the internal structure of the proton, is the light- front frame~\cite{Lorce:2018egm,Freese:2021czn}. Using both the $A(t)$ and $C(t)$ gluonic GFFs extracted from the $J/\psi-007$ experiment using the holographic QCD model, we used the following expression: 
\begin{eqnarray}
\mu_g(b) &=& M \left \{ \frac{A^{FT}_g(b)}{2} + \bar{C}^{FT}_g(b) + \frac{1}{4M^2}\frac{1}{b}\frac{d}{db} \left [ \frac{B^{FT}_g(b)}{2}- 4C^{FT}_g(b)\right ] \right \} \\
\sigma_g(b) &=& M \left \{ -\bar{C}^{FT}_g(b) + \frac{1}{2M^2}\frac{1}{b}\frac{d}{db} \left [ b \frac{dC^{FT}_g(b)}{db}  \right ] \right \}
\end{eqnarray}
 to calculate the gluon contribution to the mass density and pressure and their corresponding experimental uncertainties as a function of  the transverse position relative to the direction of motion of the proton, namely $b=r_{\perp}$. The superscript $FT$ on the form factors refers to Fourier transform of each form factor. The results are shown in Fig.~\ref{fig:ft-mup}. The quark contribution parameters are the same as those used in ref.~\cite{Lorce:2018egm} for illustration only. It is clear that the gluons mass density distribution dominates the central region of the nucleon (left panel of ~\ref{fig:ft-mup}). It is also worth noting that unlike in the Breit frame, in the light-front frame the total gluon pressure is attractive at all transverse distances balanced by the quark pressure which is mainly repulsive at almost all transverse distances. The sum of quarks anf gluons pressure, however, creates a repulsive pressure at the core mainly due to the quarks and an attractive pressure mainly due to gluons.  
 
 In conclusion, it is worth mentioning that the statistical precision of these results will be  dramatically improved using the SoLID~\cite{JeffersonLabSoLID:2022iod} detector in Hall A at Jefferson Lab. Furthermore, the universal aspect of the extracted gluons GFFs in the threshold region will ultimately be confirmed or refuted at the EIC using  $J/\psi$ electroproduction at large center of mass energies and wide range of $t$ and the other through the electro- and photo-production of $\Upsilon$ near-threshold~\cite{AbdulKhalek:2021gbh,Gryniuk:2020mlh}.

\section{Acknowledgment}
 This work is supported in part by the department of Energy Office of Science, Office of Nuclear Physics under contract DE-AC02-06CH11357.

\end{document}